\documentclass[conference]{IEEEtran}

\usepackage{cite}
\usepackage{amsmath,amssymb,amsfonts}
\usepackage{algorithmic}
\usepackage{graphicx}
\usepackage{textcomp}
\usepackage{xcolor}
\def\BibTeX{{\rm B\kern-.05em{\sc i\kern-.025em b}\kern-.08em
    T\kern-.1667em\lower.7ex\hbox{E}\kern-.125emX}}

\usepackage{xspace}
\usepackage{flushend}
\usepackage[normalem]{ulem}
\usepackage{array}
\usepackage{multirow}
\usepackage{tikz}

\long\def\ignore#1{}
\sloppypar

\newcommand*\circledred[1]{\tikz[baseline=(char.base)]{
  \node[shape=circle,draw,fill=red,text=white,font=\bf,inner sep=0.7pt] (char)
  {\scriptsize#1};
}}

\newcommand*\circledcyan[1]{\tikz[baseline=(char.base)]{
  \node[shape=circle,draw,fill=cyan,text=white,font=\bf,inner sep=0.7pt] (char)
  {\scriptsize#1};
}}

\newcommand{\etal}{\textit{et al.}}
\newcommand{\ApproxSign}{\raise.17ex\hbox{$\scriptstyle\sim$}}

\newcommand{\putsec}[2]{\vspace{-0.0in}\section{#2}\label{sec:#1}\vspace{-0.0in}}
\newcommand{\putssec}[2]{\vspace{-0.0in}\subsection{#2}\label{ssec:#1}\vspace{-0.0in}}
\newcommand{\putsssec}[2]{\vspace{-0.0in}\subsubsection{#2}\label{sssec:#1}\vspace{-0.0in}}

\newcommand{\tabref}[1]{Table~\ref{#1}}
\newcommand{\figref}[1]{Figure~\ref{#1}}
\newcommand{\secref}[1]{Section~\ref{sec:#1}}
\newcommand{\ssecref}[1]{Section~\ref{ssec:#1}}
\newcommand{\sssecref}[1]{Section~\ref{sssec:#1}}

\newcommand{\PNAME}{\mbox{FinGraV}\xspace}
\newcommand{\PNAMEBOLD}{\mbox{\textbf{FinGraV}}\xspace}

\begin{document}

\title{\PNAME:
Methodology for Fine-Grain GPU \\Power Visibility and Insights
}

\author{
    \IEEEauthorblockN{Varsha Singhania}
    \IEEEauthorblockA{
    \textit{Advanced Micro Devices, Inc.}\\
    varsha.singhania@amd.com}
    
    \and
    
    \IEEEauthorblockN{Shaizeen Aga}
    \IEEEauthorblockA{
    \textit{Advanced Micro Devices, Inc.}\\
    shaizeen.aga@amd.com}
    
    \and
    
    \IEEEauthorblockN{Mohamed Assem Ibrahim}
    \IEEEauthorblockA{
    \textit{Advanced Micro Devices, Inc.}\\
    mohamed1.ibrahim@amd.com}
}

\maketitle

\begin{abstract}
Ubiquity of AI makes optimizing GPU power a priority as large GPU-based clusters are often employed to train and serve AI models. An important first step in optimizing GPU power consumption is high-fidelity and fine-grain power measurement of key AI computations on GPUs. To this end, we observe that as GPUs get more powerful, the resulting sub-millisecond to millisecond executions make fine-grain power analysis challenging. In this work, we first carefully identify the challenges in obtaining fine-grain GPU power profiles. To address these challenges, we devise \PNAME methodology where we employ execution time binning, careful CPU-GPU time synchronization, and power profile differentiation to collect fine-grain GPU power profiles across prominent AI computations and across a spectrum of scenarios. Using the said \PNAME power profiles, we provide both, guidance on accurate power measurement and, in-depth view of power consumption on state-of-the-art AMD Instinct\textsuperscript{\texttrademark} MI300X. For the former, we highlight a methodology for power differentiation across executions. For the latter, we make several observations pertaining to GPU sub-component power consumption and GPU power proportionality across different scenarios. We believe that \PNAME unlocks both an accurate and a deeper view of power consumption of GPUs and opens up avenues for power optimization of these ubiquitous accelerators.
\end{abstract}


\putsec{intro}{Introduction}

A key enabler for the current AI wave has been the compute and memory horsepower availed by modern GPUs. As the demand for AI sky-rockets, hyperscalers continue to deploy large-scale GPU clusters to train and serve AI models~\cite{meta6kGPU,msft10KGPU}. As such, optimizing GPU power and resultant energy consumption can go a long way in reducing energy consumption and costs of AI deployments. Further, GPU power optimization is not only beneficial for AI but also for other domains, as the use cases for which GPUs are being utilized continue to widen. As an example, even for the HPC domain, the majority (more than 75\%) of node-level power provisioning for Hewlett Packard Enterprise Frontier~\cite{ornl-frontier}, the world's first exascale supercomputer, is for GPUs. 

An important first step in optimizing GPU power is better visibility into GPU power. Specifically, in this work, we focus on measuring \textit{fine-grain} GPU power profiles along both time and space dimensions. That is, first, for fine-grain profiling along time, we focus on power measurements throughout the execution of a GPU computation (commonly referred to as a \textit{kernel} in GPU parlance). Second, for fine-grain profiling along space dimension, we focus on breaking down GPU power consumption into sub-components (e.g., compute cores, memory, etc.). 

Note that, we focus on accurate kernel-level power profiles for several reasons. First, as our work and prior work~\cite{msft_polca} show, large-scale ML compute kernels can often be GPU power limited. That is, hitting GPU power limit during kernel execution causes frequency throttling~\cite{cal23} leading to lower performance. Further, as applications are comprised of sequences of kernels, power-limited kernels essentially lead to lower performance at the application-level. Finally, as energy is simply power integrated over time, accurate fine-grain power profiles lead to accurate energy measurements at the application-level aiding in energy optimization as well. Overall, we believe that fine-grain power visibility is critical in identifying avenues for, and hence design of, intelligent GPU power management capabilities.

\begin{figure}[t]
    \centering
    \includegraphics[width=0.9\linewidth]{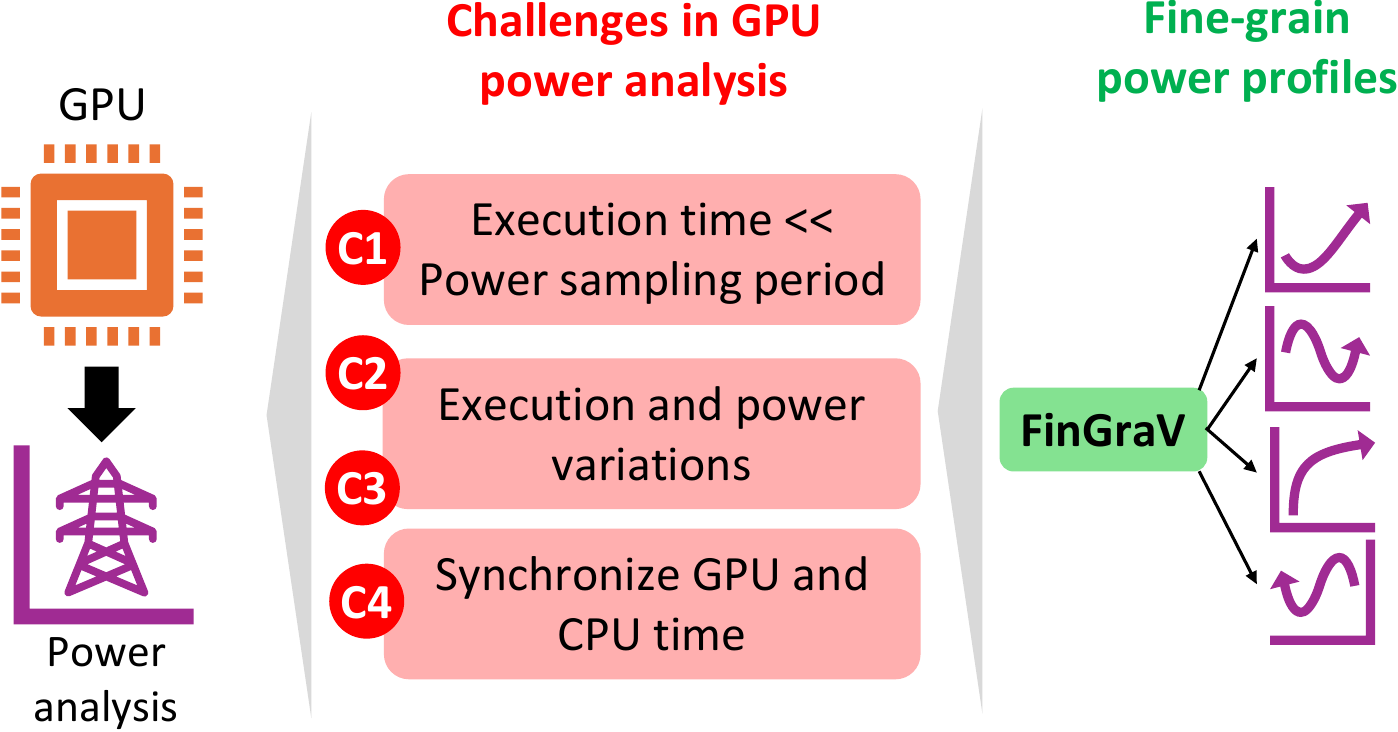}
    \caption{\PNAME addresses challenges in fine-grain GPU power analysis.}
    \label{fig:intro_overview}
\end{figure}

However, fine-grain GPU power analysis is challenging for many reasons as depicted in \figref{fig:intro_overview}. First, as GPUs get more powerful (higher compute/memory throughputs), execution times for computations can often end up in the sub-millisecond (ms) to few ms range. This makes getting fine-grain power profiles challenging as natively available power samplers on GPUs often sample power at tens of milliseconds~\cite{SCnvSMI24} (C1). 
Further, a common technique of repeated kernel executions is challenging to use as kernels with such low execution times can manifest execution time variation (e.g., due to slight differences in memory allocation and hence access patterns) which must be tackled (C2). Not only do execution time variations need to be accounted for, even power variations can occur across executions, across interleaving of various kinds of computations and more (C3) depending on averaging employed by underlying power samplers. Finally, using a high frequency GPU-side power sampler can either lead to repeated CPU-GPU communication for online measurements (as CPU schedules computations on the GPU today) or require careful synchronization of CPU-GPU time during post-processing of power logs obtained to identify logs that belong to the kernel execution (C4).

To tackle these challenges and provide fine-grain GPU power visibility, in this work, we propose \PNAMEBOLD {(\textbf{Fin}e-\textbf{Gra}in \textbf{V}isibility) methodology to collect fine-grain power profiles on the state-of-the-art AMD Instinct\textsuperscript{\texttrademark} MI300X GPU, which is being used exclusively to serve all Llama 405B live traffic at Meta~\cite{mi300x-praise}. To design \PNAME, we harness GPU-side power logging and via careful synchronization of CPU-GPU time, identify power logs of interest to stitch together a fine-grain power profile in time. Further, we employ execution time binning strategy to tackle execution time variation. Finally, we provide guidance on tackling power variation that manifests across kernel executions by aiding the user in identifying the power profile of interest for a given kernel.
Note that, as our work shows, it is critical for researchers to be aware of the said power variations as it can result in inferring inaccurate power/energy measurements (as high as 80\% energy measurement error).

We study the profiles that \PNAME leads to for prominent AI computations (matrix-matrix multiplication and collective communication kernels) across a spectrum of scenarios. Using a spectrum of these profiles, we make several observations pertaining to GPU sub-component power consumption and GPU power proportionality via comparative analysis of different AI computations. We conclude with a discussion of power optimization opportunities that \PNAME profiles help identify.

Overall, we make the following contributions in this work: 

\begin{itemize}

\item We observe that as accelerators such as GPUs get more powerful leading to lower execution times (sub-ms to few ms), their power analysis gets increasingly challenging. To this end, we first carefully identify these challenges and their implications. 

\item To address the above identified challenges, we propose \PNAMEBOLD  (\textbf{Fin}e-\textbf{Gra}in \textbf{V}isibility) methodology, where we employ execution time binning, careful CPU-GPU time synchronization, and power profile differentiation to collect fine-grain and accurate GPU power profiles.

\item Next, we employ \PNAME methodology to profile 
prominent AI computations, namely matrix-matrix multiplication (GEMM) and communication, across a spectrum of scenarios (GEMMs: compute versus memory-bound; communication: latency versus bandwidth-bound) and setups (isolated executions, interleaved executions, and more). Based on our analysis, we provide heuristics to guide fine-grain power profiling of other kernels. 

\item From the above \PNAME profiles, we make several observations pertaining to GPU sub-component power consumption and GPU power proportionality via comparative analysis of different AI computations. 

\item As GPUs continually scale compute and memory throughput, power constraints will increasingly limit them. We believe \PNAME stands to provide the necessary visibility into these ubiquitous accelerators. Based on our detailed profiling, we conclude with a discussion of opportunities to optimize power for GPUs.

\end{itemize}
\putsec{bckg}{Background}

\begin{figure}[t]
    \centering
    \includegraphics[width=0.9\linewidth]
    {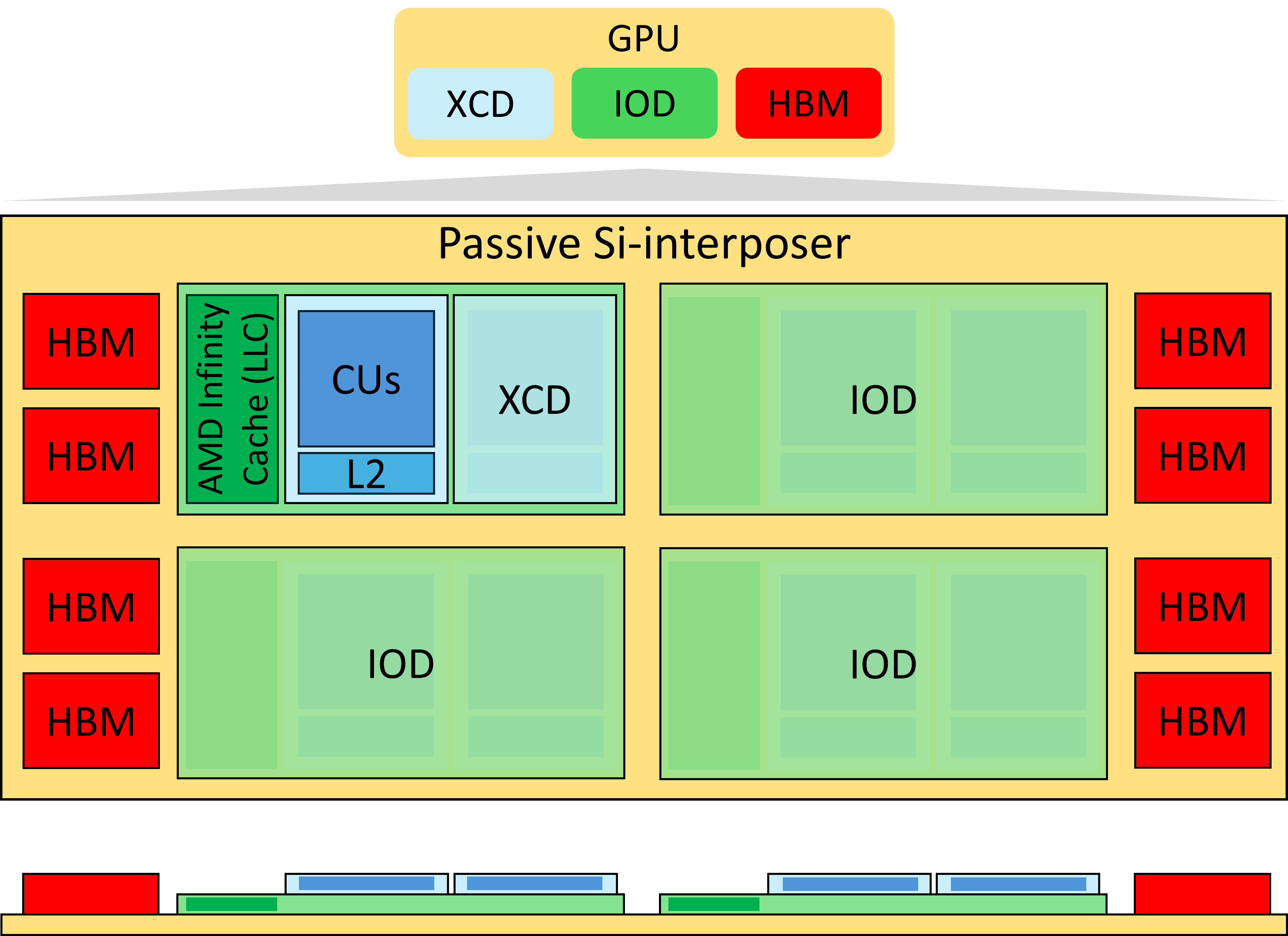}
    \caption{An illustration (not to scale) of AMD Instinct\textsuperscript{\texttrademark} MI300X, the GPU used in this work. The cross-sectional view (bottom) shows the stacking.}
    \label{fig:bckg_mi300x}
\end{figure}

\putssec{bckg_mi300x}{AMD Instinct\textsuperscript{\texttrademark} MI300X GPU and Sub-components}
In this work, we focus on fine-grain power analysis of the state-of-the-art AMD Instinct\textsuperscript{\texttrademark} MI300X GPU, based on AMD CDNA\textsuperscript{\texttrademark} 3 architecture, shown in \figref{fig:bckg_mi300x}. As depicted, the MI300X GPU is a chiplet-based design which harnesses advanced packaging to integrate heterogeneous chiplets, each specialized for a specific function. Reading the figure from the top, MI300X has accelerator complex dies (XCD) vertically stacked over I/O dies (IOD), which in turn are stacked over a passive silicon interposer. There are a total of four IODs each having two XCDs stacked for a total of eight XCDs in a single MI300X GPU~\cite{mi300x-issca,mi300x-vlsi}. 

\begin{figure*}[t]
    \centering
    \includegraphics[width=\textwidth]{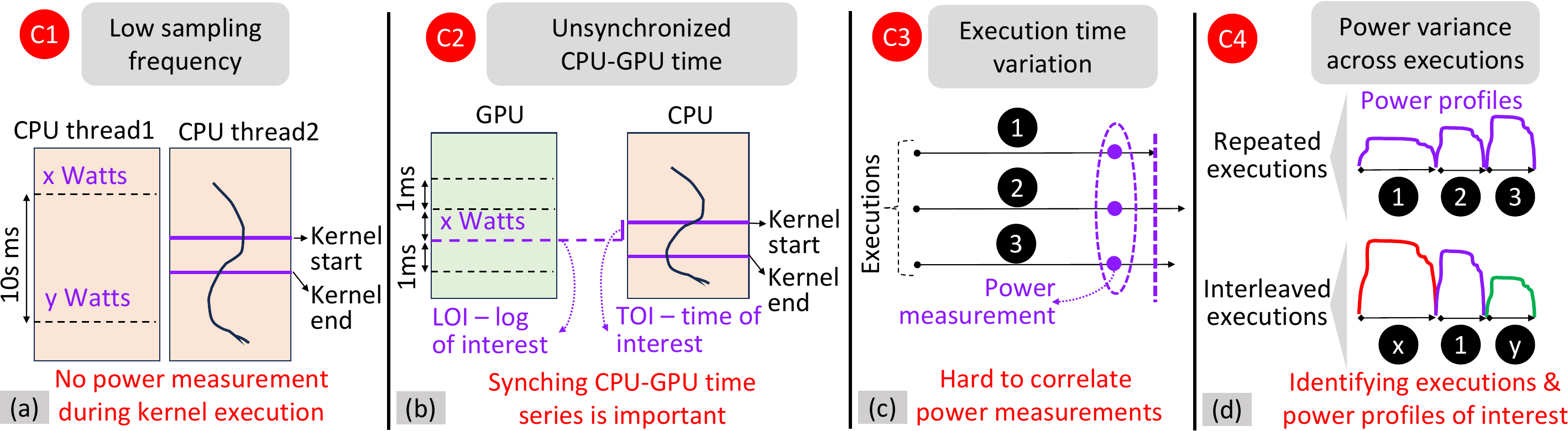}
    \caption{Challenges in doing fine-grain GPU power analysis.}
    \label{fig:challenges}
\end{figure*}

The IODs contain AMD Infinity Cache\textsuperscript{\texttrademark}, a shared memory-side last-level cache (LLC) with a total capacity of 256MB. The IODs also contain the memory interface to the on-package eight stacks of high-bandwidth memory (HBM). The total HBM capacity is 192GB (24GB per HBM stack) for a combined peak memory bandwidth of 5.3TB/s~\cite{mi300x}. The XCDs are the key computational workhorses in MI300X and each XCD in turn comprises 38 active compute units (CUs or GPU cores) for a total of 304 CUs in a single MI300X. 
The CUs within a single XCD share an L2 cache of 4MB (combined capacity of 32MB over eight XCDs). Computation is launched on GPUs in the form of a \textit{kernel} and sub-units of a kernel (termed \textit{workgroups}) are spread over available XCDs.

For large-scale AI workloads, multiple MI300X are often employed. 
In our work, we focus on the AMD MI300X Infinity Platform consisting of an 8$\times$ MI300X node with a fully-connected topology. 
That is, each MI300X is connected to seven other MI300X via 4\textsuperscript{th} Gen Infinity Fabric\textsuperscript{\texttrademark} links with a uni-directional bandwidth of 64GB/s per link~\cite{mi300x}.

\putssec{bckg_ml}{ML Focus and Operators of Interest}

While continued improvements in high-end GPUs have widened their use from high-performance computing (HPC) to AI and more, in this work, we focus on the usage of GPUs in the AI domain. We do so as the continued AI demand, and hence its ubiquity, has led to a concomitant increase in AI energy/power expenditure making it a great case study for power profiling and optimizations. As an example, Amazon's training of a 200B AI model over 48 days consumed about 11.9 GWhr~\cite{amazon200B}, equivalent to the average power consumption of over 1100 US households for a year~\cite{usHouseElectricity}. That said, our proposed methodologies will equally apply to GPU kernels in other domains. 

While AI workloads comprise a variety of operators, we consciously focus on two primary operators for our analysis, namely, general matrix-matrix multiplications (GEMM) kernels and communication kernels for they contribute to the majority of the AI execution time~\cite{patiToTC23}. We consider both GEMM and communication kernels across a spectrum of scenarios (GEMMs: compute versus memory-bound; communication: latency versus bandwidth-bound) in order to study their effect on power consumption. 
\putsec{challenges}{Challenges to Fine-grain GPU Power Analysis}

Recall that the focus of our work is to get fine-grain power profiles for GPU kernels. That is, power measurements throughout the execution of a GPU kernel (fine-grain in time dimension). Additionally, should the tooling support it, power measurement at GPU sub-component granularity (fine-grain in space dimension) such as power consumption of XCD, IOD, etc. in \figref{fig:bckg_mi300x}. However, such fine-grain GPU power analysis in time dimension is challenging due to the following.

\circledred{C1} 
\textbf{Low sampling frequency:} 
As compute throughput and memory bandwidth made available by high-end GPUs continue to scale, short of problem sizes scaling commensurately, GPU kernel execution times can often end up in the sub-millisecond (ms) to few ms range. 
This is certainly true for the AI kernels we have benchmarked on MI300X GPU (\secref{profiles}). 
This means that any GPU power measurements at low sampling frequency, as \figref{fig:challenges}a depicts, can completely miss sampling power for a given kernel. 

\circledred{C2} 
\textbf{CPU-GPU time synchronization:} 
To partially overcome the above challenge, a high sampling frequency power sampler can be used. For example, on MI300X GPU, we can tap into a 1ms power logger. However, kernel scheduling events are controlled/triggered by the CPU. 
As such, correlating this GPU power logger (agnostic of kernel start/end) with CPU time is necessary to accurately capture the power log-of-interest (LOI) and time-of-interest (TOI) as depicted in \figref{fig:challenges}b. 

\circledred{C3} 
\textbf{Execution time variation:} 
Even with the above challenges addressed, with sub-ms kernel executions, a 1ms sampler will at best deliver a single power measurement requiring multiple runs to build a fine-grain power profile. However, in the sub-ms execution space, even slight variation in kernel execution time (e.g., due to slight differences in memory allocation and hence access patterns) makes correlating power measurements across runs a challenge. As depicted in \figref{fig:challenges}c, power measurements during three separate runs are potentially at different TOI in the kernel. 

\circledred{C4} 
\textbf{Power variance across executions:} 
Finally, in addition to above execution time variation, underlying power loggers can employ averaging wherein multiple instantaneous power samples are averaged and reported at some time granularity. This in turn leads to power variations manifesting as well, which, if not tackled, can lead to considerable error in power and hence energy measurements. Specifically, as depicted in \figref{fig:challenges}d, we observe in this work that repeated executions of the same kernel (tagged 1/2/3 in \figref{fig:challenges}d) or interleaving of a kernel with other kernels (kernel 1 interleaved with kernel x and y) can lead to different power profiles. As such, identifying which power profiles to focus on is important. 

\putsec{fingrav}{\PNAME Methodology}

\begin{figure*}[t]
    \centering
    \includegraphics[width=\textwidth]{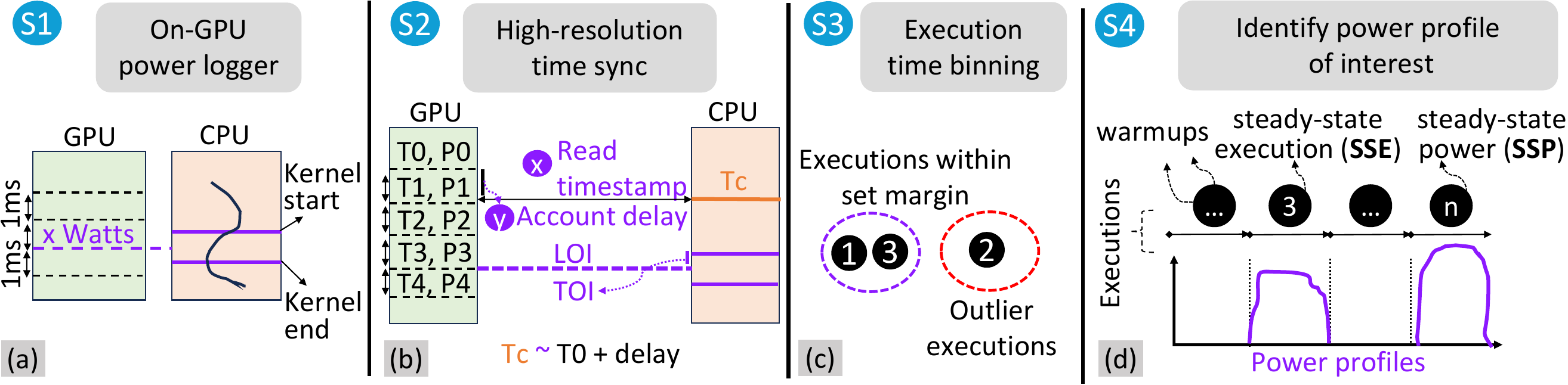}
    \caption{\PNAME strategies to address challenges in fine-grain GPU power analysis.}
    \label{fig:meth_solution}
\end{figure*}

\putssec{fingrav_solution}{Addressing Challenges}

In this section, we discuss the broad strokes of our solution to address the challenges we identified in \secref{challenges}. 

\circledcyan{S1} 
\textbf{On-GPU power logger:}  
We harness a 1ms power logger available \textbf{internally at AMD} on MI300X; each power sample is the average of multiple instantaneous power readings in the last 1ms.
We discuss using the proposed \PNAME methodology in tandem with externally available power logging tools on AMD GPUs such as \textit{amd-smi}~\cite{amdsmi} in \secref{discussion}. 
Note that, due to the averaging nature of the available power logger, the profiles in this work provide a fine-grain time view of average power manifested at different points in the kernel execution.

\circledcyan{S2} 
\textbf{High-resolution CPU-GPU sync:} 
As the internal power logger logs power measurements on the GPU while being agnostic of kernel start/end events as discussed in \secref{challenges}, we need to employ careful syncing of CPU-GPU time to identify the power log which was taken during the execution of the kernel (LOI) and where in the kernel execution was the log taken (TOI). To do so, first, we read a GPU timestamp counter before the kernel execution from the CPU side. Second, we separately benchmark the delay to read this timestamp. Finally, along with power logging, we also log this timestamp value (depicted in \figref{fig:meth_solution}b). By associating the GPU timestamp with every power log and syncing a GPU timestamp (T0) with a specific timestamp on the CPU (Tc), we can post-process the power logs to identify LOI/TOI by identifying the kernel-start/end times (also in CPU time domain) in relation to the synced CPU time Tc. 

\circledcyan{S3} 
\textbf{Kernel execution time binning:} 
Note that, as discussed in \secref{challenges}, a single run is insufficient to create fine-grain power profiles. That said, correlating measurements across runs is challenging due to the kernel execution time variation. To tackle this, we employ a simple strategy of execution time binning as depicted in \figref{fig:meth_solution}c. That is, based on our empirical experiments, we guide the user towards two heuristics for fine-grain power profiles: (1) \#runs to execute and (2) margin of execution time variation to allow. By excluding outlier runs with considerable variations and fine-tuning the margin for kernel execution time (\ssecref{profile_fingrav}), we lower the effects of execution time variation. Note that, outlier executions are important to study and while we focus on the common case in this work, we discuss power analysis for outlier executions in \secref{discussion}.

\circledcyan{S4} 
\textbf{Power profile differentiation:} 
To tackle power variations (\figref{fig:challenges}d), we provide a methodology to differentiate power profiles in order to identify the most pertinent power profile for a kernel. That is, as the underlying power logger averages instantaneous power samples in the last 1ms, power gradually rises as we continually execute a given kernel. This happens because at the start, more idle power samples are averaged with the kernel's power draw. However, with repeated executions, as the kernel executions fill up power averaging window of 1ms, a time series view of average power of the kernel emerges and power does not vary substantially beyond this point. 

Consequently, we identify two specific power profiles as depicted in (\figref{fig:meth_solution}d). First, we tag the power profile of the first kernel execution, post warm-up executions, beyond which the kernel execution time does not lower substantially (typically three warm-up executions from GPU idle state), as the steady-state execution (SSE) profile. Without power profile differentiation as we suggest, this is the power profile that a typical user associates with a kernel. Additionally, we tag the power profile of the kernel execution post SSE, beyond which the power does not vary substantially as the steady-state power (SSP) profile. Note that, as power depends on voltage, frequency and temperature, SSP profile is by definition specific to a given voltage-frequency setting and as such can be affected by dynamic voltage-frequency scaling. That said, in this work, we did not control the decisions of the underlying power management infrastructure.

It is the SSP power profile that provides the time series view of average power at different points in kernel execution. Comparing SSP and SSE profiles, demonstrates the potential error in power (and hence energy) measurement that can manifest without the power profile differentiation that we recommend. Finally, note that, depending on how large the kernel execution time is in comparison to the power averaging time window (1ms for our setup), SSP and SSE profile can be the same.

\putssec{fingrav_steps}{\PNAME: Steps}

Bringing together the solutions identified above, we list the steps we follow in \PNAME methodology:

\begin{enumerate}
    \item Time the kernel five times to identify the \textbf{kernel execution time}. Use this to lookup the \PNAME empirical guidance table (\tabref{tab:finGraV_guidance}) to deduce the recommended \#runs, binning margin, and a guidance on \#LOIs to collect for the given kernel execution time. 
    
    \item Add relevant \textbf{CPU-side instrumentation} to GPU code. 
    This includes timing the kernel start/end, reading the GPU timestamp before kernel execution (\ssecref{fingrav_solution}), and starting/ending power logger before/after the kernel. 
    
    \item For the \textbf{SSE} profile only, per run, execute the kernel four times. Note that, across all AI computations we studied, three executions sufficed for execution time stabilization. That said, this can be simply deduced for a given GPU/library/binary empirically.
    
    \item For the \textbf{SSP} profile, ascertain the number of executions needed using: max (ceil [(averaging interval)\slash(kernel execution time)], executions for SSE), where the averaging interval for our setup is 1ms and executions for SSE are four. Note that an SSP run can also get the SSE profile. Note also that, should power (frequency) throttling incur during warmup runs (rise followed by fall of power), \textbf{binary search} can be necessary to deduce executions to get SSP profile.
       
    \item \textbf{Execute} specified runs with specified executions per run. Note that, to attain power averages at unique TOI in kernel (\figref{fig:challenges}b), we add random delays before starting kernel executions per run. 
    
    \item Discard all but the \textbf{golden runs}. Golden runs are those that include SSP execution times belonging to the execution time bin with the maximum number of executions within binning margin of each other (for example 5\% as shown in \tabref{tab:finGraV_guidance} for 25-200us entries). 
    
    \item \textbf{Synchronize CPU-GPU time} to identify LOI, if available, in the power logs collected per run. 
    Identify also the TOI (the time in kernel execution where LOI was obtained).
    
    \item If \#LOIs obtained so far are less than those suggested in \tabref{tab:finGraV_guidance}, \textbf{optionally, execute more runs} (= \#LOIs). 
    
    \item \textbf{Stitch the different runs} by plotting all collected LOIs and TOIs.
\end{enumerate}

\begin{table}[t]
    \centering
     \caption{\PNAME Profiling Guidance.}
    \label{tab:finGraV_guidance}
    \begin{tabular}{|l|l|l|l|l|}
     \hline
        Exec & \# Runs & \# LOI & Binning\\
        range & & & margin\\ \hline
        25-50us         & 400  & 1/5us   & 5\%  \\ \hline
        50-200us        & 200  & 1/10us   & 5\%  \\ \hline
        200us-1ms       & 200  & 1/10us   & 2\%  \\ \hline
        \textgreater1ms & 200  & 1/10us   & 2\%  \\ \hline
    \end{tabular}
    \vspace{-\baselineskip}
\end{table}

\putsec{profiles}{\PNAME Profiles and Insights}

We begin with a discussion of AI computations under study in this work and our setup for executing them. We follow this by providing an evaluation of the key tenets of \PNAME methodology and sharing some experimental profiling guidance. Finally, we discuss \PNAME profiles for AI computations, key observations from these profiles, and implications for future hardware/software based on these observations. 

\putssec{profiling_ai_setup}{AI Operator Space and Setup}

As discussed in \ssecref{bckg_ml}, in this work, we focus on two primary operators for our analysis, namely, general matrix-matrix multiplication (or GEMM, M$\times$K * K$\times$N = M$\times$N) kernels and communication kernels occurring in AI workloads as they contribute to the majority of AI execution time~\cite{patiToTC23}. 

Specifically, we cover compute-bound (CB) square (M=N=K) GEMM sizes of (8K=8192, 4K=4096, 2K=2048) and memory-bound (MB) GEMV sizes for the same matrices (i.e., M=K, N=1) for a total of six AI GEMMs. We define a kernel to be compute-bound if its algorithmic op-to-byte ratio is larger than the machine's op-to-byte as calculated from the peak compute and memory throughput of the underlying processor (kernel is memory-bound otherwise). Additionally, for communication, we study multi-GPU collectives such as all-gather and all-reduce which are widely used in AI workloads. For collectives, we consider both latency-bound (64KB and 128KB, relevant for inference) and bandwidth-bound (512MB and 1GB, relevant for training) scenarios. Note that collective kernels, depending on associated data-transfer size, can be latency or bandwidth-bound. We classify a size as latency-bound if collective latency at/before this size does not increase commensurate to data-transfer size (kernel is bandwidth-bound otherwise). 

To execute GEMMs, we harness AMD ROCm\textsuperscript{\texttrademark}~\cite{rocm} rocBLAS library~\cite{rocblas} (version 4.2.0) consisting of high-performance GEMM kernels. For AI collectives, we employ AMD ROCm\textsuperscript{\texttrademark} Communication Collectives Library (RCCL)~\cite{rccl} (version 2.20.5), a library of standard collective communication routines for GPUs. As discussed in \ssecref{fingrav_steps}, in a given run, we execute a kernel multiple times and collect power logs over the entire run. We use post-processing to identify executions of interest within a run. 

\putssec{profile_fingrav}{\PNAME Methodology Evaluation}

Before we present AI power profiles, we first begin with an evaluation of the key tenets of \PNAME methodology namely: (a) the benefit of CPU-GPU time sync, (b) power profile differentiation benefit, (c) the effect of kernel execution time binning, and (d) resiliency to \#runs executed. To do so, we focus on multiple power profiles for a compute-bound (CB) 4K GEMM kernel (hence referred to as CB-4K-GEMM for simplicity) depicted in \figref{fig:fingrav_meth} with and without these techniques. In the figure, we depict multiple runs of a CB-4K-GEMM and multiple executions within a run.
We have time for a run on the x-axis and total power profiled on the y-axis.\footnote{In this work, we focus on relative power data (not absolute numbers).}\footnote{The total power, on the y-axis, is the voltage regulator output power.}
Additionally, using vertical line markers, we separate the warmup, SSE (steady-state execution) and SSP (steady-state power) executions on the graph (\ssecref{fingrav_solution}). 

\begin{figure}[t]
    \centering
    \includegraphics[scale=0.2]{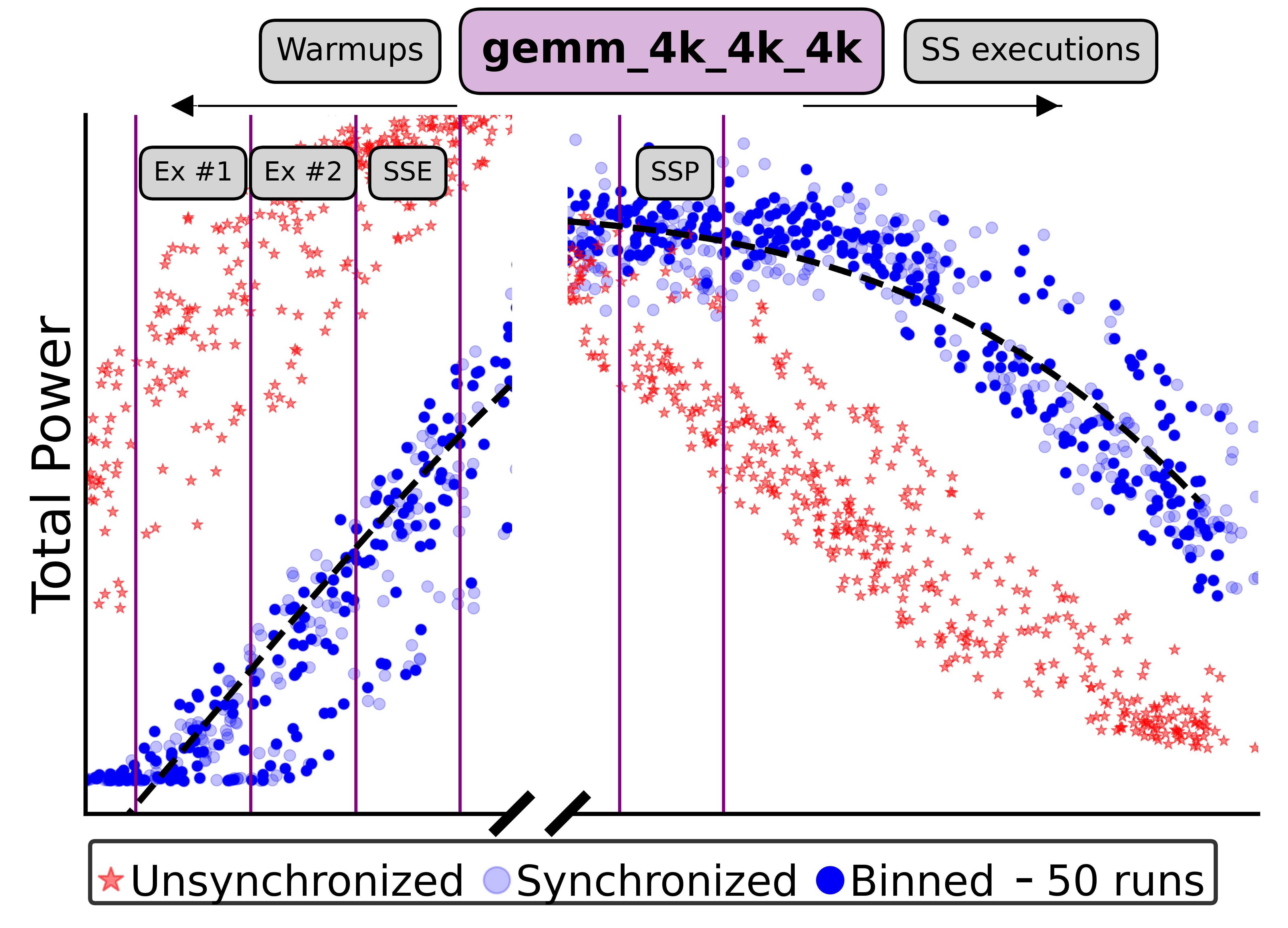}
    \caption{\PNAME methodology evaluation for (a) benefit of CPU-GPU time sync, (b) effect of kernel execution time binning, and (c) resiliency to \#runs using CB-4K-GEMM power profiles under different scenarios.}
    \label{fig:fingrav_meth}
\end{figure}

\noindent\textbf{Benefits of CPU-GPU Time Sync:} 
We discussed in \ssecref{fingrav_solution} the importance of syncing CPU-GPU time while using a GPU-based power logger for events triggered by the CPU. \figref{fig:fingrav_meth} shows the benefit of this time synchronization for CB-4K-GEMM by comparing the unsynchronized (red) power profile to the synchronized (blue) power profile. As shown, the synchronized profile captures the gradual rise in power as GPU moves from idle state to executing the kernel (warm-up executions, to SSE, then finally to SSP) while the unsynchronized profile misses this ramp and fails to align power changes with appropriate executions in a run.

\noindent\textbf{Accurate Power Profiles with Power Profile Differentiation:} 
\figref{fig:fingrav_meth} shows that SSE and SSP profiles differ considerably. Consequently, assuming SSE profile as the kernel's power profile can lead to up to 36\% error in power (and hence energy) measurement. 

\noindent\textbf{Benefits of Kernel Execution Time Binning:} 
We also discussed, in \ssecref{fingrav_solution}, the importance of kernel execution time binning to better tackle execution time variation. We show how this binning leads to tighter power profiles in \figref{fig:fingrav_meth} where we show the profile without binning using transparent blue dots while the profile with binning is shown with filled/dark blue dots. As shown, binning leads to a tighter power profile, more tuned to the true shape of power consumed. Tighter binning margins can even further smoothen the power profile albeit at the cost of more \#runs as we discuss below with the profiling guidance we offer.

\noindent\textbf{Resiliency to Executed \#Runs:} 
We discussed in \ssecref{fingrav_solution} that with a 1ms power logger and sub-ms kernels, we get at best a single power log in a given run. We need multiple runs to create fine-grain power profiles. While in subsequent graphs we indicate a certain \#runs, we discuss here, the effect of considerably lowering the \#runs on the power profile. All the CB-4K-GEMM power profiles in \figref{fig:fingrav_meth} use about 200 runs but we also depict (with a dashed black line) the power profile that can be attained with 50 runs only. To get this line, we do a linear regression of degree four over the power data we get with 50 runs only. As shown, even with 50 runs, we are able to ascertain the overall power trend for CB-4K-GEMM. 
Consequently, while in the rest of the paper we use large \#runs for smooth power profiles, fewer \#runs can also be employed. 

\noindent\textbf{\PNAME Profiling Guidance:}
Finally, we show in \tabref{tab:finGraV_guidance} some profiling guidance largely driven by our empirical analysis for GEMM kernels. The table covers the ranges of executions we see for GEMM kernels. We observed that for smaller kernel execution times, more \#runs and slightly higher kernel binning margin can be needed to get enough power LOIs to create smooth power profiles. That said, this is simply a guidance and, as we discussed above, \PNAME methodology is resilient to lowering \#runs. 
\putssec{gemm}{GEMM Profiles and Insights}

\begin{figure}[t]
    \centering
    \includegraphics[scale=0.7]{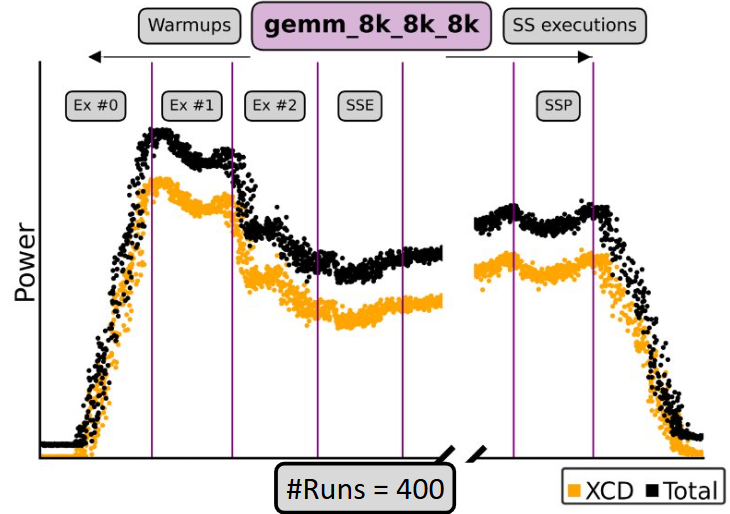}
    \caption{CB-8K-GEMM total and XCD power.}
    \label{fig:CB_8K_GEMM}
\end{figure}

We discuss our GEMM power profiles (compute-bound/CB and memory-bound/MB) and learnings in this section. We start with general power trends we observe, follow that with component-level (\ssecref{bckg_mi300x}) comparative analysis, and finally discuss power behavior in the presence of interleaved GEMM executions. Along the way, we make several key observations, provide guidance for accurate power measurements and make recommendations for future hardware/software to optimize GPU power (all summarized in \tabref{tab:finGraV_insights}).

\putsssec{gemm_trends}{Power Trends}

\begin{figure*}[t]
    \centering
    \includegraphics[scale=0.49]
    {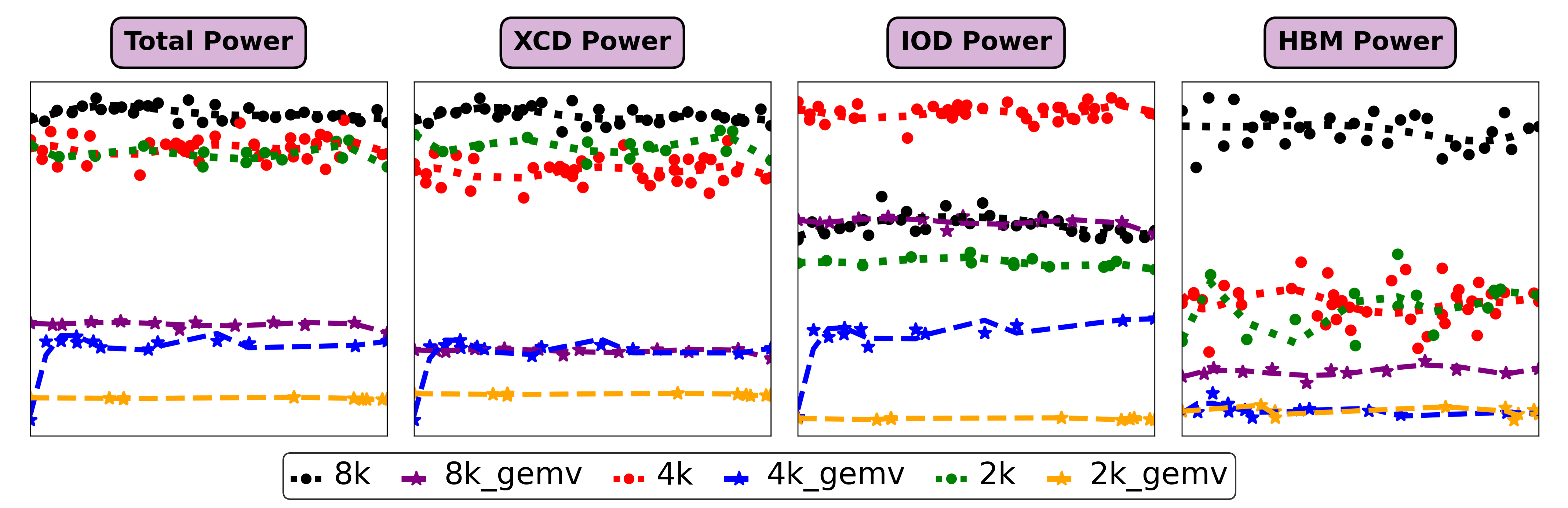}
    \caption{Component-level comparative analysis of compute-bound GEMMs and memory-bound GEMVs.}
    \label{fig:gemm_comparison}
\end{figure*}

We first begin with the general question of power trends manifested. To do so, we depict the power profile (total and XCD component) for 8K compute-bound GEMM (CB-8K-GEMM) over 200 runs with multiple executions per run in \figref{fig:CB_8K_GEMM}. We make several observations here. First, the power rises for initial executions and then it drops till it reaches SSE profile. Finally, the power slightly increases till it reaches SSP profile at the last execution after which it shows little variation (\ssecref{fingrav_solution}).

We observe that for a compute-heavy GEMM kernel such as CB-8K-GEMM (high op-to-byte ratio), the first few executions considerably stress power, invoking the power management firmware to throttle frequency~\cite{cal23} in order to manage power excursions. We also observe that the best execution time, which happens for SSE execution and beyond, shows lower power than these initial executions. Finally, the SSP execution shows slightly higher power than the SSE execution. 

We compare these observations to the power profile for CB-2K-GEMM (\figref{fig:CB_2K_GEMM}) which is considerably less compute-heavy in comparison to CB-8K-GEMM (based on op-to-byte ratio). We observe that power starts low for initial executions before rising considerably for the SSP execution.
Recall that all executions between SSE and SSP have similar execution times. 

\begin{figure}[t]
    \centering
    \includegraphics[scale=0.55]{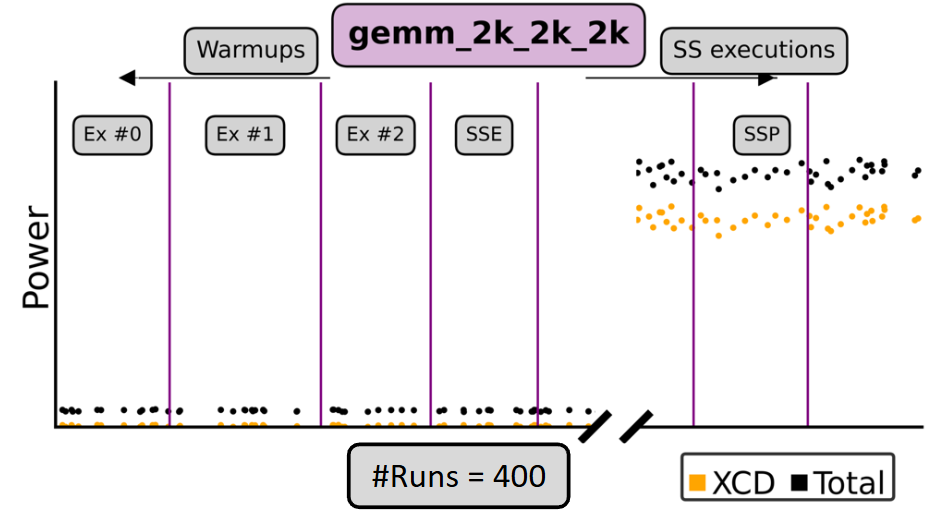}
    \caption{CB-2K-GEMM total and XCD power.}
    \label{fig:CB_2K_GEMM}
    \vspace{-\baselineskip}
\end{figure}

Based on above two power profiles, overall, the two key power trends we observe are (a) a sharp rise followed by a drop to a steady-state and (b) a gradual rise to a steady-state. This provides credence to \PNAME power profile differentiation strategy that eventually power stabilizes giving a time-series view of average power at different points in kernel execution time (SSP profile). Further, as execution time of CB-8K-GEMM is longer than power averaging window (1ms) while CB-2K-GEMM is much shorter, we observe that the resultant spread between SSE and SSP power, and hence power/energy measurement error, is much higher for CB-2K-GEMM (80\% vs. 20\% for CB-8K-GEMM).

The above observations lead us to our first key takeaway (highlighted in \tabref{tab:finGraV_insights}) that similar kernel execution times (SSE through SSP) can manifest very different power profiles. Further, not factoring this variation can lead to a measurement error of as high as 80\% depending on relative magnitudes of kernel execution time and power averaging time window of underlying power logger. This also leads to our first power measurement guidance that power profile differentiation (SSE, SSP) as employed by \PNAME is very crucial for accurate power measurements.

\putsssec{gemm_components}{Component Comparative Analysis}

Next, we compare component-level power across all three CB GEMMs and three MB GEMVs. \figref{fig:gemm_comparison} depicts, in a relative manner, the total and component-level powers (XCD, IOD, HBM). Note that we use SSP power profiles of these kernels to plot these graphs for as discussed above SSP profile represents the true time-series view of average power for different points in kernel execution. Finally, for better visibility, we also show linear regression lines for all power profiles. 

We make several observations here. All CB GEMMs show considerably higher total and XCD power versus MB GEMVs. This makes sense as CB GEMMs incur heavy compute and also data movement, while MB GEMVs are data movement heavy only.\footnote{As we repeatedly execute kernels, data movement is heavily biased toward on-chip data movement for our executions.} Amongst CB GEMMs, CB-8K-GEMM has slightly higher total/XCD power. On the GEMV front, we see a drop in total power going from 8K-GEMV to 2K-GEMV. Unlike total/XCD power, where CB GEMMs dominate, MB-8K-GEMV does stress IOD power. Finally, as discussed above, as our data movement is biased towards on-chip data movement (repeated executions), both CB/MB GEMM kernels stress HBM power similarly. 
The exception is CB-8K-GEMM, which has the highest HBM power, as its considerably large input sizes stress the on-chip caches the most.

\begin{table*}[t]
\centering
\caption{\PNAME Profiling Insights, Power Measurement Guidance and, Recommendations for Future Hardware/Software.}
\label{tab:finGraV_insights}
\resizebox{\textwidth}{!}{%
\begin{tabular}{|m{0.01\textwidth}|m{0.38\textwidth}|m{0.44\textwidth}|m{0.09\textwidth}|}
\hline
\multicolumn{1}{|c|}{\textbf{\#}} & \multicolumn{1}{c|}{\textbf{Takeaway}} & \multicolumn{1}{c|}{\textbf{Power Measurement Guidance/Recommendation}} & \multicolumn{1}{c|}{\textbf{Section}} \\ \hline
1 & Similar kernel execution times can manifest very different power profiles depending on relative magnitudes of kernel execution time and power averaging time-window of underlying power logger. & \textbf{Measurement guidance-1}: \PNAME power profile differentiation is crucial to deduce a kernel's power profile without which high measurement errors (as high as 80\%) can manifest. & \sssecref{gemm_trends} \\ \hline
2 & Total power scales with work done and different GPU components get stressed based on algorithmic nature of underlying computation. & \textbf{Recommendation-1}: Available power headroom can be fully utilized by concurrently executing computations with complementary algorithmic and hence complementary power profiles. & \sssecref{gemm_components} \\ \hline
3 & Compute-heavy kernels are dominated by XCD component power. &  \textbf{Recommendation-2}: Techniques that optimize XCD component power should be prioritized to optimize overall total power for compute-heavy kernels. & \sssecref{gemm_components} \\ \hline
4 & Compute-light kernels and compute-heavy kernels show similar XCD component power. & \textbf{Recommendation-3}: Techniques to attain GPU power proportionality are necessary for compute-light kernels especially for XCD component. & \sssecref{gemm_components} \\ \hline
5 & Power for certain kernels (memory-bound GEMVs and compute-light GEMMs) gets affected by kernels preceding them while for other kernels (compute-heavy GEMMs) it is not affected. & \textbf{Measurement guidance-2}: Isolated executions are necessary to assess kernel's power draw when its execution time is shorter than power averaging time-window of underlying power logger. & \sssecref{gemm_interleaved} \\ \hline
\end{tabular}%
\vspace{-\baselineskip}
}
\end{table*}

The above observations lead us to our second key takeaway (\tabref{tab:finGraV_insights}) that total power generally scales with work done (e.g., CB \textgreater~MB) with different components getting stressed based on the underlying algorithm for a given computation (e.g., CB stress XCD power, MB can stress IOD power, etc.). That is, algorithmic behavior (op-to-byte) is, to a first order, a good indicator of component-level GPU power consumption. This takeaway in turn leads to our first recommendation that, should it be possible, we should stitch together kernels with complementary power profiles at an algorithmic/software level. This, with enough power headroom, will allow us to reap the benefits of concurrent execution. An example of this for AI workloads is concurrent execution of MB attention kernel with CB fully-connected layers~\cite{zhu2024nanoflow}.

Further note that, combining the data in \figref{fig:CB_8K_GEMM} and \figref{fig:CB_2K_GEMM}, it is clear that total power and XCD power are very close to each other for CB GEMMs. This, together with low IOD/HBM power depicted in \figref{fig:gemm_comparison}, leads to our third key takeaway that XCD component is the most dominant component for CB GEMMs which leads, by Amdahl's law, to our second recommendation that techniques that focus on optimizing XCD power are crucial to optimize overall power for CB computations.

Another takeaway we point to is that all CB GEMMs are in the ballpark of each other when it comes to XCD power. That said, compute throughput calculated using algorithmic ops and execution times shows that CB-2K-GEMM has about half the compute utilization in comparison to CB-4K/8K-GEMM. Note that, unlike XCD power, we observe that the IOD power tracks well with LLC bandwidth. This leads us to our third recommendation that GPU power proportionality, that is, expending power commensurate to rate of work, needs further investigation especially with a focus on XCD component for compute-light kernels (e.g., CB-2K-GEMM with comparatively lower compute utilization). Of particular focus can be different software execution schedules which are performance-iso but lead to different (hopefully lower) power profiles.

\begin{figure}[t]
    \centering
    \includegraphics[scale=0.36]{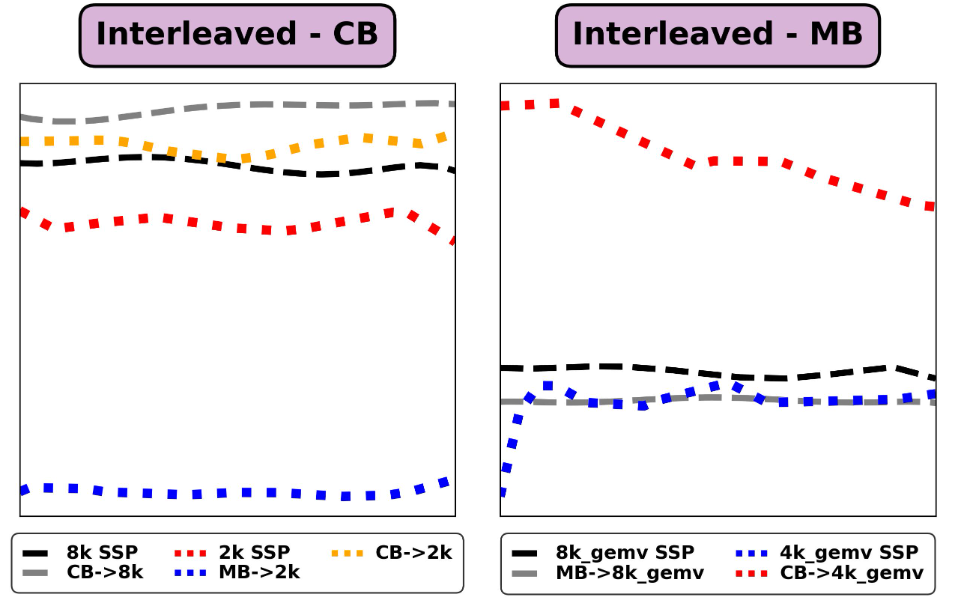}
    \caption{Total power comparison for different interleaved GEMM and GEMV.}
    \label{fig:gemm_interleaved}
\end{figure}

\begin{figure*}[t]
    \centering
    \includegraphics[scale=0.49]
    {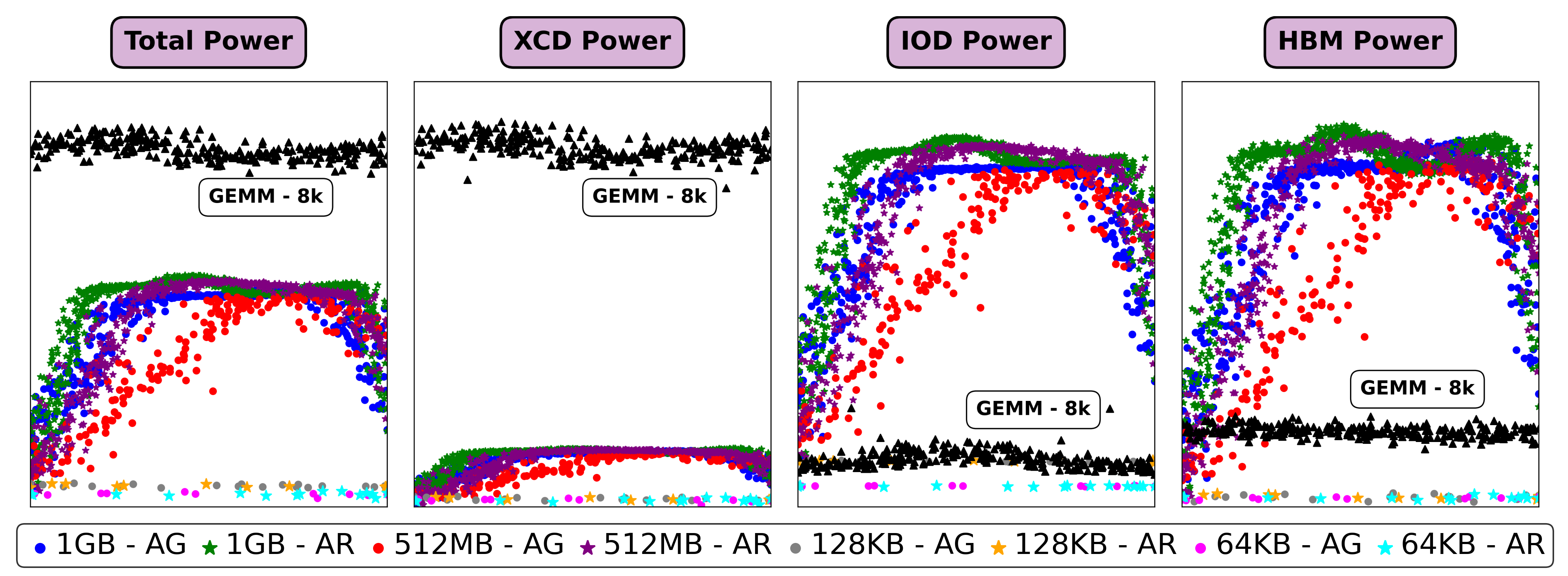}
    \caption{Component-level comparative analysis of the evaluated communication kernels and CB-8K-GEMM.}
    \label{fig:comm_comparison}
\end{figure*}

\putsssec{gemm_interleaved}{Interleaved Kernels Analysis}

Next, we aim to tease out how the power profiles change when different kinds of kernels are executed in an interleaved fashion. To do so, we plot relative total powers for a given GEMM/GEMV and compare the resultant power profile to its SSP power in isolation. 
This is depicted in \figref{fig:gemm_interleaved}. 

Focusing on the left graph, we first compare the SSP profile of CB-8K-GEMM, a compute-heavy kernel, to its profile when it is run post 60 CB-2K-GEMMs (CB--\textgreater 8K). We observe a slight rise in CB-8K-GEMM power in relation to its SSP power. Next, we compare the SSP profile of CB-2K-GEMM, a compute-light kernel, across different interleaved executions and see that there is a considerable difference in the observed power profiles. When 40 MB-4K-GEMVs are run before a single CB-2K-GEMM (MB--\textgreater 2K), the power for CB-2K-GEMM is far lower than its SSP profile. In comparison, if we run CB-8K-GEMM and CB-4K-GEMM before CB-2K-GEMM (CB--\textgreater 2K), its power is higher than its SSP power. This shows that the power of the compute-light kernels is affected by which kernels precede them but that of a compute-heavy kernel is not.

We see something similar for MB kernels in the graph on the right in \figref{fig:gemm_interleaved}. MB kernels are also affected by kernels that precede them. Comparing the SSP profile for MB-8K-GEMV, when other MB kernels (MB-4K/2K-GEMV) are interleaved before it, shows that MB-8K-GEMV consumes less power in comparison to its  SSP profile (MB--\textgreater 8K\_gemv). On the other hand, MB-4K-GEMV shows more than its SSP power when CB kernels (CB-8K/4K-GEMM) are run before it (CB--\textgreater 4K\_gemv).

The above observations lead us to our fifth key takeaway that certain kernels (MB GEMVs and compute-light GEMMs) are affected by kernels preceding them when it comes to their power consumption. The primary reason for this behavior is that the execution time of these kernels is considerably shorter than power averaging time window of the power logger we use. As such, their measured power profiles are an average of power manifested by preceding kernels and the kernel of interest. This in turn leads to our second measurement guidance that for kernels whose execution time is lower than power averaging time-window, isolated executions are necessary to assess the true kernel power draw. Note that, with an instantaneous power sampler, \PNAME methodology can assess power for kernels regardless of their execution time and regardless of run setup (isolated or interleaved executions).

\putssec{comm}{Communication Profiles and Insights}

We next focus on power analysis for communication kernels. Recall that we profile two widely used AI communication kernels: all-gather (AG) and all-reduce (AR). 
Further, we exercise both latency-bound/LB (64KB/128KB) and bandwidth-bound/BB scenarios (512MB/1GB) for both kernels. \figref{fig:comm_comparison} presents relative powers (total, XCD, IOD, and HBM) for all eight communication kernels. We also plot CB-8K-GEMM in the same graph for comparing GEMM and communication kernel power profiles. 

We make several observations here. With regards to XCD power, CB-8K-GEMM has much higher power than communication kernels, which is expected. That said, when it comes to the total power, BB communication kernels fall somewhere in the middle of LB communication and CB-GEMM kernels. This can be explained by the considerably higher IOD and HBM power incurred by BB communication kernels. 

Based on the comparison of communication and computation-heavy (GEMM) kernels, our takeaway is similar to takeaway \#2 in \tabref{tab:finGraV_insights} and consequently our resultant recommendation is also the same in that the heterogeneous power profiles manifested can be exploited for efficient concurrent execution should there be enough power headroom (e.g., latency-bound communication in parallel with any other computation, etc.). 
\putsec{discussion}{Discussion}

\noindent\textbf{\PNAME with External Power loggers:} 
In our work, we harness a power logger available internally at AMD on MI300X that provides average power samples for multiple instantaneous power readings in last 1ms (\ssecref{fingrav_solution}). We believe that the general \PNAME  methodology we have can be applied to external power loggers available on AMD platforms such as amd-smi~\cite{amdsmi}. The key tenets of \PNAME (careful CPU-GPU time synchronization, kernel execution time binning, and power profile differentiation) are all equally relevant even with these external power loggers. All of this said, the resultant power profiles will heavily depend on the power information these external loggers report. 
As an example, since these loggers may report average power over longer time-windows, any such averaging done can impact the power profiles that \PNAME methodology produces. We leave further augmenting \PNAME for these tools to future work. 

We believe that, in addition to \PNAME methodology, a key contribution of our work is indeed the power profiles for key AI operations of GEMMs and communication kernels that we make available. These profiles provide both an accurate and deeper (different components such as XCD, IOD, HBM) view of power consumption on a state-of-the-art MI300X GPU which led to some targeted recommendations for power optimizations (\tabref{tab:finGraV_insights}). Finally, we also provided power measurement guidance (\tabref{tab:finGraV_insights}), which will aid researchers in avoiding common pitfalls and steep errors that can happen with power (and hence energy) measurements on high-end GPUs (as high as 80\% error). All of these are immensely valuable in designing power-optimized future accelerators.

\noindent\textbf{Outlier Executions:} 
We discussed in \ssecref{fingrav_solution} that we focus \PNAME profiles on the common kernel execution time and discard any outliers. While we believe that prioritizing power analysis for the common case is the right first step, understanding the power behavior of the outlier executions is also important. One way to attain \PNAME profiles for outlier executions is to employ \PNAME methodology and focus on collecting profiles for a specific outlier execution time and discarding the rest (that is changing step-6 in \ssecref{fingrav_steps}). While doable, this can be costly as more \#runs can be necessary to create a large enough bin belonging to this specific outlier execution time. 
Another strategy could be to break down kernel execution into phases when possible and assess if this lowers variation in each phase (as compared to variation at the kernel-level). As an example, with GPU kernels, wherein each kernel launches multiple workgroups, the kernel can be artificially terminated after half the number of workgroups are completed and each half of the execution can be studied separately. We leave such investigations to future work.  
\putsec{related}{Related Work}

\noindent\textbf{Power/Energy Measurement Methodologies:}
Prior work measured GPU power using vendor tools and interfaces~\cite{amdsmi,nvsmi} to access native on-board telemetry readings. Tools, such as Variorum~\cite{variorum}, harness these interfaces to provide a vendor-neutral library for power measurement across different hardware from multiple vendors. Such efforts to standardize the power measurement interface, along with exposing more telemetry and control, are critical for exascale systems as highlighted by national labs and hyperscalers~\cite{sandia_standard,msft_polca}.
To sample power at a higher rate compared to vendor tools, other work physically measure power using external power meters~\cite{pwr_meter}. Using these meters, prior work assessed the existing on-board power measurement and found discrepancies in the reported power, highlighting the importance of using a high-fidelity methodology for measuring and reporting power~\cite{K20_2014,cao2020_arxiv,Shahid_2020,Jay_2023,SCnvSMI24}. 
Concurrent to our work, Yang \etal{}~\cite{SCnvSMI24} did a comprehensive analysis of non-AMD GPUs and identified similar power measurement guidance as we do (e.g., skip/shift mechanism they employ is equivalent to SSE/SSP power profile differentiation we employ). However, they do not evaluate AMD GPUs and further, they do not focus on power optimization recommendations. Finally, researchers built simulators and statistical models to provide fine-grain power and energy estimates~\cite{accelwattch,gpujoule,adhinarayanan_iiswc16,IPDPS14,CCGrid16,HPCA15}. 

In this work, we focus on native on-board power measurements, instead of modeling, as the GPUs are getting more complex, pushing both compute and memory throughputs and hence power.
With its fine-grain power profiles, \PNAME can be used to improve the fidelity of the power models.      
As discussed in \secref{discussion}, \PNAME can work with existing public power loggers (with the caveats discussed) and can complement prior power measurement methods by providing a step-by-step methodology to unlock fine-grain visibility for the evaluated kernels.
This way \PNAME can be agnostic to the varying sampling rate of the power measurement interfaces and meters. 
As the kernel executions get shorter, \PNAME key principles will be more critical to collect sane fine-grain power profiles. 

Lang \etal{}~\cite{europar13} aimed to construct high-resolution power profiles similar to \PNAME.
However, \PNAME addresses emerging trends and challenges in GPUs that prior work ignored such as execution time variation of short kernels (\secref{challenges}).
Unlike Lang \etal{}, \PNAME addresses this variation using kernel execution time binning. 
Also, Lang \etal{} used repetitive CPU-GPU synchronization to address the drift between CPU and GPU clocks over time. However, the authors did not factor in the delays imposed by the CPU-GPU communication.
We observed such drift and will address this challenge in future work. 

\noindent\textbf{Power/Energy Characterization and Optimization:}
Researchers used the tools and methods above to characterize the energy efficiency of critical workloads and primitives in AI and HPC running on different scales~\cite{Price_2015,CCGrid20,adamek_access_2021,jahanshahi_cal20,patterson2021carbon,white2022,yu_hpca23}, and to study the efficiency of the latest innovations in GPUs and other accelerators~\cite{cal23,ispass24}. 
Prior work also investigated the impact of frequency capping, power capping, DVFS, and input data composition on energy efficiency~\cite{Mei_2013workshop,WORKS19,eenergy19,Krzywaniak_2020,msft_polca,gregersen_arxiv24}. 

With the above characterization, researchers found optimization opportunities to improve the performance-per-watt within a single GPU and in large-scale deployments. 
Per-GPU optimizations focused on tuning DVFS policies during kernel execution to balance energy efficiency and performance ~\cite{DynPowerHPCA2017,Srikant_2022,Zhang_ECCS_2024,WangDRLCAP2024}, while other work additionally investigated the impact of tuning workload parameters~\cite{gogreen_2022,JayaweeraCGO24}. 
As for system-level optimizations, recent work identified and reduced the energy bloat during training~\cite{ZeusYou23,perseusChung24,envpipe_usenix23} and inference~\cite{DynamoLLM24} by controlling job (e.g., batch size, server instances, and model parallelism) and GPU (e.g., frequency and power cap) knobs. 
Other work focused on efficient power management in large-scale LLM inference by enabling power oversubscription in LLM clusters~\cite{msft_polca} or via deploying the different phases of LLM (prompt and token generation) on different machines in the cluster~\cite{splitwise24}.   

We harnessed \PNAME to collect fine-grain power profiles for sub-ms GEMM/V kernels and communication collectives, key AI primitives, running on AMD MI300X. 
These power profiles, across the sub-components of MI300X, unveiled insights related to the different power behaviors of the evaluated kernels. 
Using \PNAME fine-grain profiles, researchers can uncover opportunities to design better GPU power managers to improve the energy efficiency of both, standalone GPUs and large-scale GPU clusters.   
\putsec{conclusion}{Conclusion}

We focus in this work on fine-grain profiling for key AI operations of matrix-matrix multiplication and communication kernels on GPUs, which are widely deployed for AI workloads. To this end, we first identified challenges in doing fine-grain GPU power profiling and proposed \PNAME methodology to address these challenges on a state-of-the-art AMD MI300X GPU. Our work identifies important power measurement guidance to avoid steep measurement errors (as steep as 80\%) in power (and hence energy). Additionally, we identify several important takeaways from \PNAME power profiles which in turn lead to power optimization recommendations focused on GPU sub-component power consumption and GPU power proportionality.

\section*{Acknowledgment}
The authors thank Nuwan Jayasena, Ashish Jain, Steve Kushnir, Aranyak Mishra, Yuri Lee, and the anonymous ISPASS reviewers for helping improve the paper. 
AMD, the AMD Arrow logo, AMD CDNA, AMD Instinct, AMD ROCm, AMD Infinity Cache, AMD Infinity Fabric, and combinations thereof are trademarks of Advanced Micro Devices, Inc. Other product names used in this publication are for identification purposes only and may be trademarks of their respective companies.


\bibliographystyle{IEEEtran}
\bibliography{ref}

\end{document}